\documentclass[12pt]{article} 
\usepackage[sectionbib]{natbib}
\usepackage{array,epsfig,rotating}
\usepackage[]{hyperref}  
\usepackage{sectsty, secdot}
\usepackage{comment}
\usepackage{amsmath}
\usepackage{amsmath,bbm}
\usepackage{graphicx} 
\usepackage{enumerate}
\usepackage{url} 
\usepackage{bm}
\usepackage{booktabs}

\subsectionfont{\fontsize{12}{14pt plus.8pt minus .6pt}\selectfont}

\usepackage{amsmath}
\usepackage{amssymb}
\usepackage{amsfonts}
\usepackage{multirow}
\usepackage{amsthm}
\usepackage{algorithm}
\usepackage{algorithmic}
\usepackage{authblk}

\usepackage{caption}
\usepackage{subcaption}

\theoremstyle{definition}

\usepackage{setspace}
\usepackage{xcolor}

\title{Uncertainty Analysis of Experimental Parameters for Reducing Warpage in Injection Molding}
\author[1]{Yezhuo Li\thanks{
Part of this work is based in part on Yezhuo Li's doctoral dissertation  at Clemson University.}}
\author[2]{Fan Zhang\thanks{First two authors contributed equally to this work.}}
\author[3,4]{Dhanashree Shinde}
\author[1]{Qiong Zhang\thanks{Corresponding author:qiongz@clemson.edu}}
\author[3]{Sai Pradeep}
\author[3,4,5,6]{Srikanth Pilla}
\author[7]{Gang Li}
\affil[1]{School of Mathematical and Statistical Sciences, 
Clemson University}
\affil[2]{Department of Mathematics, Boise State University}
\affil[3]{Center for Composite Materials, University of Delaware}
\affil[4]{Department of Materials Science and Engineering, University of Delaware}
\affil[5]{Department of Mechanical Engineering, University of Delaware}
\affil[6]{Department of Chemical and Biomolecular Engineering, University of Delaware}
\affil[7]{Department of Mechanical Engineering, Clemson University}
\date{~}
\begin{document}

\maketitle
\newpage
\begin{abstract}
Injection molding is a critical manufacturing process, but controlling warpage remains a major challenge due to complex thermomechanical interactions. Simulation-based optimization is widely used to address this, yet traditional methods often overlook the uncertainty in model parameters. In this paper, we propose a data-driven framework to minimize warpage and quantify the uncertainty of optimal process settings. We employ polynomial regression models as surrogates for the injection molding simulations of a box-shaped part. By adopting a Bayesian framework, we estimate the posterior distribution of the regression coefficients. This approach allows us to generate a distribution of optimal decisions rather than a single point estimate, providing a measure of solution robustness. Furthermore, we develop a Monte Carlo-based boundary analysis method. This method constructs confidence bands for the zero-level sets of the response surfaces, helping to visualize the regions where warpage transitions between convex and concave profiles. We apply this framework to optimize four key process parameters: mold temperature, injection speed, packing pressure, and packing time. The results show that our approach finds stable process settings and clearly marks the boundaries of defects in the parameter space.
\end{abstract}


\section{Introduction}

Injection molding (IM) of plastics is a widely adopted manufacturing process due to its cost-effectiveness, design flexibility, and rapid production capabilities \citep{zhao2022recent, farahani2022data, pradeep2024novel}. It is commonly used to produce thermoplastic components in industries such as automotive, electronics, and consumer products \citep{lavaggi2022theory, zarei2022design}. The IM process typically consists of three key thermomechanical stages: filling, packing, and cooling. Due to the viscoelastic nature of thermoplastics, variations in temperature and pressure across these stages can introduce internal residual stresses that deteriorate part quality and lead to various defects.

Among these defects, warpage represents a major dimensional distortion arising from non-uniform shrinkage. This non-uniformity often results from uneven temperature gradients, excessive pressure during the packing stage, or differential stress relaxation in polymer chains during cooling \citep{mohan2017review, wang2018optimization, farahani2022data}. To mitigate warpage, numerous studies have aimed to optimize key process conditions that control the residual stress distribution and reduce warpage \citep{kuo2022simple}. 
The real experiments can be time consuming and expensive. Therefore, 
simulation-based studies play a critical role in guiding the design of real-world experiments, as they enable systematic exploration of the parameter space at relatively low cost and risk. 
However, warpage sensitivity often depends on part geometry, gating configuration, and local thickness variations.
A case-specific investigation remains necessary to  derive guidelines applicable to different product designs in practice.

\begin{figure}
 \centering
 \includegraphics[scale=0.45]{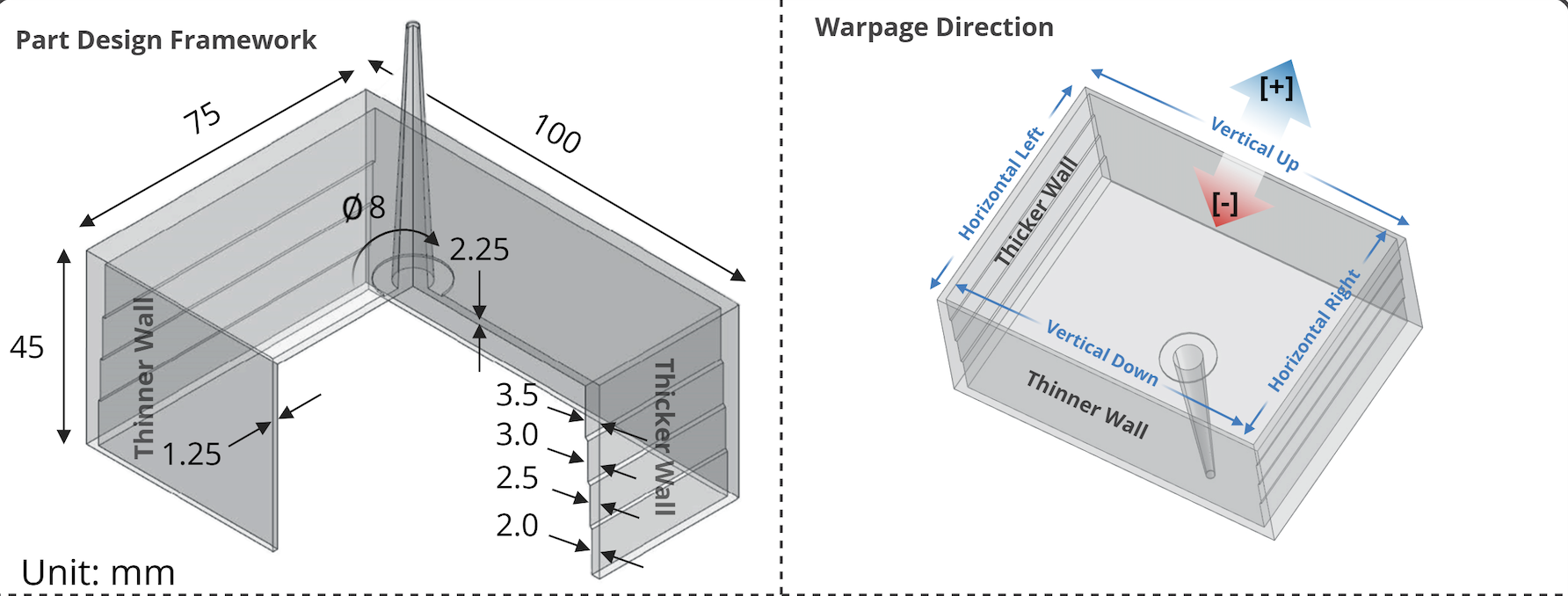}
 \caption{Left: CAD model of the test geometry highlighting dimensional details, including asymmetric wall thicknesses, 1.25 mm and 2 mm. Right: warpage direction nomenclature and sign convention is categorized as positive (convex) or negative (concave) based on surface displacement.}\label{fig:Partdesignwarpage}
 \end{figure}

In this paper, we consider a box-shaped geometry adapted from \cite{gim2024mold}, which serves as a representative model for enclosure-type components frequently used in consumer and automotive applications.
We employ the commercial injection molding simulation software Moldex3D to model warpage behavior under four influential process parameters: mold temperature, injection speed, packing pressure, and packing time. 
In the left panel of Figure \ref{fig:Partdesignwarpage}, we illustrate the detailed part design of a box-shaped geometry, which has dimensions of 100 mm (width) × 75 mm (length) × 45 mm (depth). This part design features a central sprue gate and incorporates two distinct wall thickness configurations. The shorter opposing side walls, each 2 mm thick, are designated as horizontal left and horizontal right. A stepped transition from 3.5 mm to 2 mm is included along one edge to introduce localized geometric variation. The longer side walls, referred to as vertical up and vertical down, have a reduced thickness of 1.25 mm.

As shown in the right panel of Figure \ref{fig:Partdesignwarpage}, the warpage refers to the  surface displacement along the positive (convex) or negative (concave) direction for all four walls. The ideal situation is that the displacement metrics of the four walls are all close to zero. The scientific question is to find the parameter setting that can reduce the overall warpage. Therefore, our objective is to minimize the sum of squared  displacement metrics across the four walls
\[
(displacement~on~horizontal~left)^2+
(displacement~on~vertical~down)^2+
\]
\begin{equation}\label{eq:objective}
(displacement~on~horizontal~right)^2+
(displacement~on~vertical~up)^2,
\end{equation}
with respect to the process parameters-mold temperature, injection speed, packing pressure, and packing time.
In addition, a related scientific objective is to characterize the boundaries in the parameter space that distinguish positive (convex) from negative (concave) displacement regions for each wall.

The simulation of the IM process does not have a closed-form expression. Therefore, we cannot directly assess the optimal parameter setting for the objective in \eqref{eq:objective}. In this paper,
we solve the two scientific questions
from a data driven perspective. 
Given a set of data with inputs and outputs collected from running the simulation, we fit polynomial regression models to approximate the displacement metrics of four walls, respectively. 
By adapting a Bayesian framework to the polynomial regression model, we provide uncertainty analysis 
for the optimal decision. Also, based on the Bayesian polynomial model, we propose a boundary analysis and visualization framework to address the second scientific question. Our approach is fundamentally aligned with the application of polynomial approximation models for experimental data, a framework that originates from the classical literature on response surface methodology (RSM) \citep{myers1996response}. 
RSM has been widely applied across diverse engineering disciplines to build predictive models and optimize systems using experimental data. For instance, manufacturing engineering uses RSM to model how process parameters like cutting speed affect quality responses such as surface roughness, helping to identify optimal operating windows \citep{NOORDIN200446}. Similarly, in chemical engineering, it is employed to numerically optimize process conditions like reaction temperature and catalyst concentration to maximize a final product yield \citep{HAMZE20151}.
Recently, the Bayesian uncertainty estimation of optimal decision was developed under the Gaussian process surrogate model in \cite{li2025uncertainty} for the application of curing process in manufacturing. However, these existing approaches are not able to address our two scientific objectives directly.

The remainder of this paper is organized as follows. Section \ref{sec2} describes the proposed approach of estimating decision uncertainty with polynomial regression model. Section \ref{sec3} implements boundary analysis to estimate decision uncertainty originated from parameter uncertainty. 
In Section \ref{sec4}, we apply the proposed method to the IM simulation experiment; Section \ref{sec5} concludes the paper.

\section{Decision Uncertainty Estimation with Polynomial Surrogate Models}
\label{sec2}
Consider a composite optimization problem
\begin{equation}\label{eq:objeq}
\bm x^\ast\in\mathrm{argmin}_{\bm x\in\mathcal X} G(\bm y(\bm x)),
\end{equation}
where $\bm x=(x_1, \ldots, x_d)^\top\in \mathcal X\in \mathbb R^d$ is the $d$ dimensional decision variable, 
\[
\bm y(\bm x)=(y_1(\bm x), \ldots, y_m(\bm x))^\top
\]
is a vector of simulation outputs which are black-box functions of the inputs $\bm x$, and $G(\cdot)$ is a function with a known closed-form expression. For the application of the warpage reduction for the injection molding process, $\bm x$ is a four dimensional vector containing the experimental parameters mold temperature, injection speed, packing pressure and packing time, $\bm y(\bm x)$ is a four dimensional vector containing the displacement metrics of the four walls in Figure \ref{fig:Partdesignwarpage},
and the objective function $G(\cdot)$ is the squared sum of the displacements 
\begin{equation}\label{eq:imobj}
G(\bm y(\bm x))=\sum^m_{l=1}y^2_l(\bm x)
\end{equation}
as illustrated in \eqref{eq:objective}.

Our goal is to assess the optimal decision based on a limited budget of simulation runs, and also provide uncertainty estimation of the optimal decision. We use polynomial regression models to surrogate each simulation output, respectively. Let $\mathcal D=\{\bm x_1, \ldots, \bm x_n\}$ be a set of input points, and 
$\bm y_l=(y_l(\bm x_1), \ldots, y_l(\bm x_n))^\top$ be a vector collecting the $l$-th simulation output for $l=1,\ldots, m$. We can 
express the polynomial model by
\begin{equation}\label{eq:poly}
    \bm y_l = \bm P\bm{\beta}_l + \bm \epsilon_l,\quad\mathrm{for}\quad l=1, \ldots, m
\end{equation}
where 
$\bm P$ is an $n \times p$ design matrix constructed by polynomial functions
of $\mathcal D$:
\[\bm P =\begin{bmatrix}\bm p^\top(\bm x_1),\bm p^\top(\bm x_2),\cdots,
\bm p^\top(\bm x_n)
\end{bmatrix}^\top\]
with $\bm p(\bm x)$ representing a $p$-dimensional polynomial functions of $\bm x$ including the intercept. 
The linear coefficients $\bm{\beta}_l$ is a vector of size $p$ and $\bm{\epsilon}_l=\left( \epsilon_{l1}, \cdots, \epsilon_{ln} \right)^\top$ is the error vector with i.i.d entries from the normal distribution with zero mean and variance $\sigma^2_l$. It is possible to consider the dependence of $\bm{\epsilon}_l$
across different outputs, say, for $l=1, \ldots, m$. For the injection molding application we are considering in this paper,
after removing the effects explained by the linear model,
there is no strong evidence from the data to support more complex dependence structures. Therefore, we fit polynomial regression models separately for each output without modeling the dependence of errors across different directions. 
Under the assumption of the polynomial regression model, the ordinary least squares (OLS) estimator for the coefficient vector $\bm{\beta}_l$ is given by
$\hat{\bm{\beta}}_l = \left(\bm P^\top \bm P\right)^{-1} \bm P^\top \bm y_l$
 for $l=1, \ldots, m$. An estimation of the optimal decision in \eqref{eq:objeq} is given by
 \begin{equation}\label{eq:surrogate}
 \hat{\bm x}^\ast\in \mathrm{argmin}_{\bm x\in\mathcal X} G\left[\bm p^\top(\bm x)\hat{\bm{\beta}}_1, \ldots,  \bm p^\top(\bm x)\hat{\bm{\beta}}_m\right].
 \end{equation}
 Given the form of $G()$ for the injection molding application in  \eqref{eq:imobj}, the objective is a polynomial function with respect to $\bm x$. Therefore, the surrogate optimization problem in \eqref{eq:surrogate} simplifies the original problem in \eqref{eq:objeq}.

The uncertainty of the optimal decision given by \eqref{eq:surrogate} can be estimated under the Bayesian framework of linear regression. 
We specify a non-informative prior for $\bm{\beta}_l$, such that $p(\bm{\beta}_l) \propto 1$, and an independent Jeffrey’s prior for $\sigma^2_l$, given by $p(\sigma^2_l) \propto {\sigma^{-2}_l}$. The posterior distribution of $\bm{\beta}_l$ given $\sigma^2_l$ follows a multivariate distribution
\begin{equation}\label{eq:mvn}
\bm{\beta}_l \mid \bm{y}_l, \sigma^2_l \sim \mathcal{MVN}\left(\hat{\bm{\beta}}_l, \sigma^2_l (\bm P^\top \bm P)^{-1}\right),
\end{equation}
where $\sigma^2_l$ has posterior distribution $\sigma^2_l \mid  \bm{y}_l \sim \text{Inv-Gamma}\left(\frac{n}{2}, \frac{1}{2}\|\bm{y}_l - \bm{P} \hat{\bm{\beta}_l}\|^2\right)$, and 
\begin{equation}\label{eq:sig}
\hat{\sigma}^2_l=\frac{\|\bm{y}_l - \bm{P}\hat{\bm{\beta}_l}\|^2}{n-p}
\end{equation}
is the maximum a posteriori (MAP) estimator \citep{pishro2014introduction} of $\sigma^2_l$. The posterior uncertainty of $\bm\beta_l$ lead to uncertainty in the optimal decision in \eqref{eq:objeq}. We can numerically assess this uncertainty.
By plugging in the MAP estimator in 
\eqref{eq:mvn}, we denote $\bm\beta^{(i)}_l$ for $l=1, \ldots, m$
and $i=1,\ldots, R$ as 
realizations of $\bm\beta_l$ from 
the multivariate normal distribution. 
For $i=1, \ldots, R$, we obtain 
\begin{equation}\label{eq:xuncertainty}
    \bm{x}^{(i)} \in  \mathrm{argmin}_{\bm x\in\mathcal X} G\left[\bm p^\top(\bm x)\bm{\beta}^{(i)}_1, \ldots,  \bm p^\top(\bm x)\bm{\beta}^{(i)}_m\right].
\end{equation}
The set $\{\bm{x}^{(1)}, \ldots, \bm{x}^{(R)}\}$ provides a distribution of the optimal decision, and therefore, can be used to assess decision uncertainty led by the data
under the model assumption.

\begin{figure}[!h]
    \centering
    \includegraphics[scale=0.6]{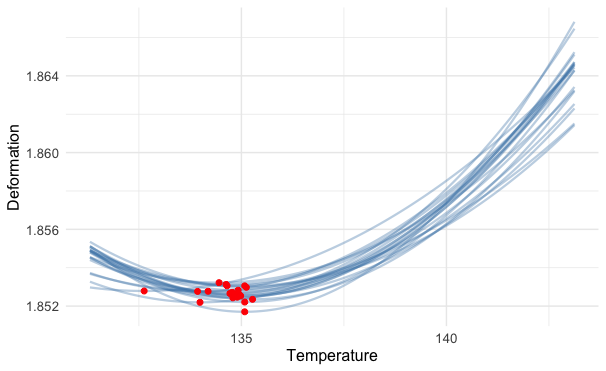}
    \caption{Multiple realizations of the quadratic function (blue lines) and their corresponding optimal decisions (red dots) for the polynomial model in \eqref{eq:cure}.}
    \label{fig:1d-polynomial}
\end{figure}

\begin{figure}[!h]
    \centering
    \includegraphics[scale=0.6]{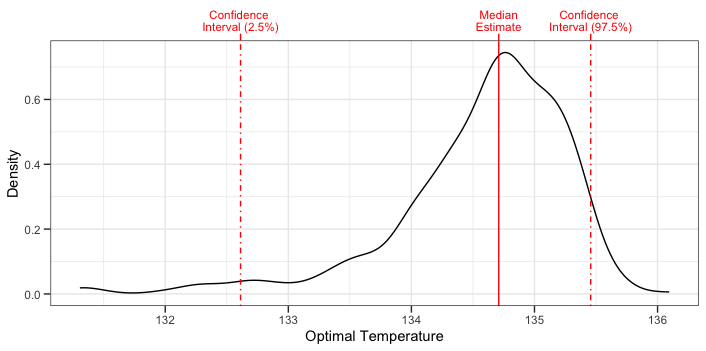}
    \caption{Empirical density of the uncertainty set in \eqref{eq:xuncertainty} for the polynomial model in \eqref{eq:cure}.}
    \label{fig:LOO}
\end{figure}

\paragraph{Illustration Example 1.}
We consider a simplified manufacturing example to illustrate this idea. This example considers the cure of thermoset-based fiber-reinforced composite laminates from \cite{li2025uncertainty}.
The optimization problem aims to minimize deformation induced by residual stresses with respect to the end temperature of the first stage, while keeping all other parameters of the cure cycle fixed.
We model the deformation $y(x)$ as a second order polynomial of the temperature $x\in \left[ T_1, T_2 \right]$:
\begin{equation}\label{eq:cure}
  y(x)  = \beta_0+\beta_1x+\beta_2 x^2+\epsilon,
\end{equation}
By fitting the linear model to simulation data from \cite{li2025uncertainty}, we obtain an uncertainty set of the optimal decision $x^\ast$ based on multiple realizations of a quadratic function, whose coefficients are drawn from the posterior distribution. The resulting uncertainty set of $x^\ast$ 
is illustrated in Figure \ref{fig:1d-polynomial}. Using this set, we estimate the probability density of the optimal temperature and construct a 95\% confidence interval as shown in Figure \ref{fig:LOO}. From a practical standpoint, these results suggest maintaining the curing temperature between 133 and 135 to minimize deformation during the curing process.

\section{Boundary Analysis with Uncertainty}
\label{sec3}

The method introduced in the previous section can be used to perform uncertainty analysis of parameter settings that 
minimizes the overall deformation. In this section, we investigate 
boundaries in the parameter space that distinguish positive from negative  displacement regions for each outcome, also construct 
confidence bands of the boundaries for uncertainty quantification.

Our objective is to find the boundary of concave and convex displacements across the parameter space for each direction, i.e.,
for $l=1,\ldots, m$, we are interested in locating the boundary
\[
\partial\mathcal X_l=\{\bm x\in \mathcal X~|~y_l(\bm x)=0\}
\]
where $\partial$ denotes the boundary of the set. We also aim to label the regions in $\mathcal X$ with
\[
\mathcal X^+_l=\{\bm x\in \mathcal X~|~y_l(\bm x)\geq 0\}\quad\mathrm{and}
\quad 
\mathcal X^-_l=\{\bm x\in \mathcal X~|~y_l(\bm x)\leq 0\}.
\]
It is also possible that the boundary does not exist in the parameter space $\mathcal X$, and $\mathcal X$ is a subset of $\mathcal X^+_l$ or $\mathcal X^-_l$.
We facilitate boundary analysis with the polynomial surrogate model in \eqref{eq:poly}. With the fitted model coefficients $\hat{\boldsymbol{\beta}_l}$, we can obtain the estimated boundary
\[
\widehat{\partial\mathcal{X}}_l=\{\bm x\in \mathcal X~|~\bm p^\top(\bm x)\hat{\boldsymbol{\beta}}_l=0\},
\]
The estimated positive and negative regions can be obtained accordingly.

To quantify the uncertainty of the boundary, we define
the $100\times (1-\alpha)\%$ confidence band of $\partial\mathcal X_l$ with error tolerance $\varepsilon>0$ as
\[
\mathcal{C}_l(\alpha,\varepsilon)=\left\{\bm x\in\mathcal X:\mathbb P\left(\left|\bm p^\top(\bm x)\bm{\beta}_l\right| < \varepsilon \big\rvert \bm y_l\right)\geq 1-\alpha\right\}
\]
given the posterior distribution of $\boldsymbol\beta_l$ in \eqref{eq:mvn}. 
For polynomial surfaces, this inversion usually results in more complex regions that are difficult to describe analytically. To address this, we adopt a Monte Carlo approach. This simulation-based method allows us to approximate the boundary uncertainty directly by sampling from the posterior distribution of the regression coefficients.
Therefore, we approximate it using realizations $\bm{\beta}^{(i)}_l$'s
from the posterior distribution of $\bm{\beta}_l$. 
Let 
\[
y^{(i)}_l(\bm{x}) = \bm{p}^\top(\bm{x}) \bm{\beta}^{(i)}_l,
\]
we can define the boundary with the realization $y^{(i)}_l(\bm{x})$
\[
\partial\mathcal X^{(i)}_l=\{\bm x\in \mathcal X~|~y^{(i)}_l(\bm{x})=0\}
\]
for $i=1, \ldots R$ as in \eqref{eq:xuncertainty}.
We approximate the confidence band 
\[
\hat{\mathcal{C}}_l(\alpha,\varepsilon) = \left\{\bm{x} \in \mathcal{X} : \left|\frac{y^{(i)}_l(\bm{x})}{\sigma^y_l(\bm{x})}\right| \leq \varepsilon \text{ for at least }(1-\alpha)  \text{of samples } i \right\},
\]
where 
\[
\sigma^y_l(\bm{x}) = \sqrt{\hat{\sigma}^2_l \bm{p}^\top(\bm{x})(\bm{P}^\top \bm{P})^{-1} \bm{p}(\bm{x})}.
\]
We use the scaling term $1/\sigma^y_l(\bm{x})$ to standardize the prediction uncertainty. In regression models, the prediction error is not the same everywhere. It usually depends on the location of $\bm{x}$. By dividing the response by its standard deviation, we convert the value into a standardized scale. The parameter $\varepsilon$ acts as a threshold. It controls the width of the confidence band in units of standard deviation. A larger $\varepsilon$ leads to a wider band, which covers more uncertainty. 

It is worth noting that the proposed method is different from 
analytical simultaneous confidence bands, such as Scheffé's method \citep{bohrer1967sharpening}. 
Those methods are used to quantify the uncertainty of the response $y(\bm x)$, whereas our goal is to characterize the uncertainty of the boundary in the input space $\bm x$. 

\paragraph{Illustration Example 2.}
We illustrate the boundary analysis approach using a synthetic example with one dimensional output and two dimensional input
$\bm x=(x_1,x_2)^\top$:
\[
y(\bm x) = -82.17 - 2.01  x_1 - 1.61  x_2 + 2.4  x_1^2 + 3.76  x_2^2 - 1.2  x_1  x_2+\epsilon,
\]
where $x_1, x_2 \in [-10, 10]$ and the error term $\epsilon$ follows a normal distribution with mean zero and variance $\sigma^2$. The variance $\sigma^2$ is drawn from a Gamma distribution with the shape parameter $\alpha=2$ and the rate parameter $\beta=1$.
We use a Latin Hypercube design \citep{loh1996latin} with 500 runs to generate inputs, and obtain the corresponding outputs. We fit the data using a second order polynomial regression model as illustrated in Section \ref{sec2}.
Based on the fitted model, we generate $R=500$ realizations of $y^{(i)}(\bm x)$ and construct the 95\% confidence band with different tolerance values $\varepsilon$. 

\begin{figure}[H]
    \centering
    \includegraphics[scale=0.55]{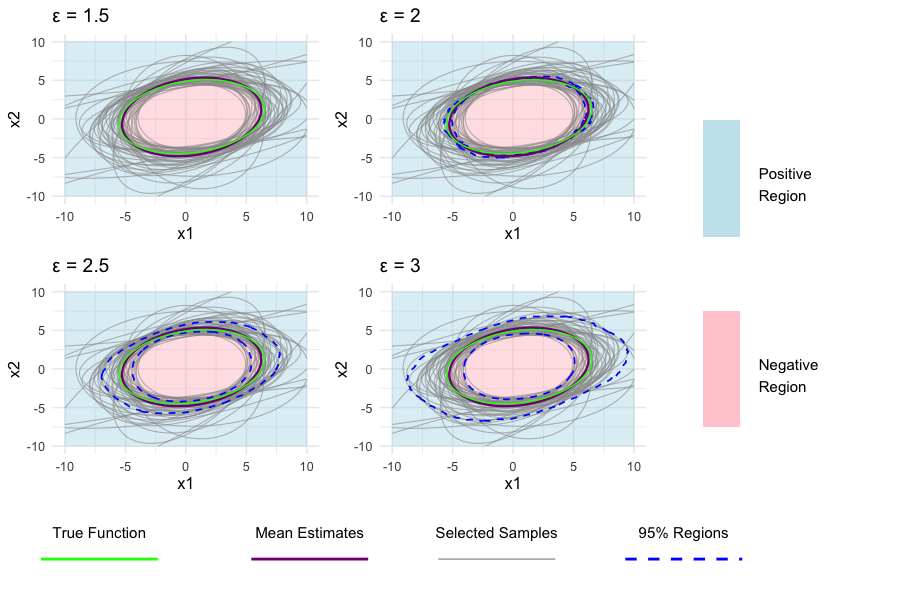}
    \caption{Panel of Monte Carlo-based confidence bands for the quadratic surrogate model in Example 2 at four standardized tolerances. Each subplot shows the 95\% boundary (blue dashed contour) approximately satisfying \(\mathbb P\bigl(|y(x_1,x_2)/\sigma_y(x_1,x_2)| < \varepsilon\bigr)\ge0.95\) with \(\varepsilon\) equal to 1.5, 2.0, 2.5, and 3.0 standard deviations, respectively. The purple solid curve is the posterior mean zero-level set, the green solid curve is the original true function zero-level set, and the gray curves are ten random posterior sample zero-level sets.}
    \label{fig:mc-band}
\end{figure}

Figure~\ref{fig:mc-band} presents a panel of four Monte Carlo-based confidence bands for the quadratic surrogate model corresponding to standardized tolerances \(\varepsilon = 1.5, 2.0, 2.5,\) and \(3.0\). In each subplot, the blue dashed contour marks the approximate confidence band $\hat{\mathcal C}(\alpha, \varepsilon)$.
The purple solid curve depicts the approximate boundary $\widehat{\partial{\mathcal X}}$, the green solid curve shows the  true boundary $\partial\mathcal X$, and the gray curves trace ten randomly selected posterior sample $\partial\mathcal X^{(i)}$.
Figure~\ref{fig:mc-band} shows how the confidence bands change with different \(\varepsilon\) values. When \(\varepsilon\) increases, the band covers a larger area. A small \(\varepsilon\) gives a narrow band and tight boundary. A large \(\varepsilon\) gives a wide band and loose boundary. Users need to adjust \(\varepsilon\) according to their requirements.

\section{Application to Injection Molding}
\label{sec4}

\begin{figure}
    \centering
    \includegraphics[scale=0.5]{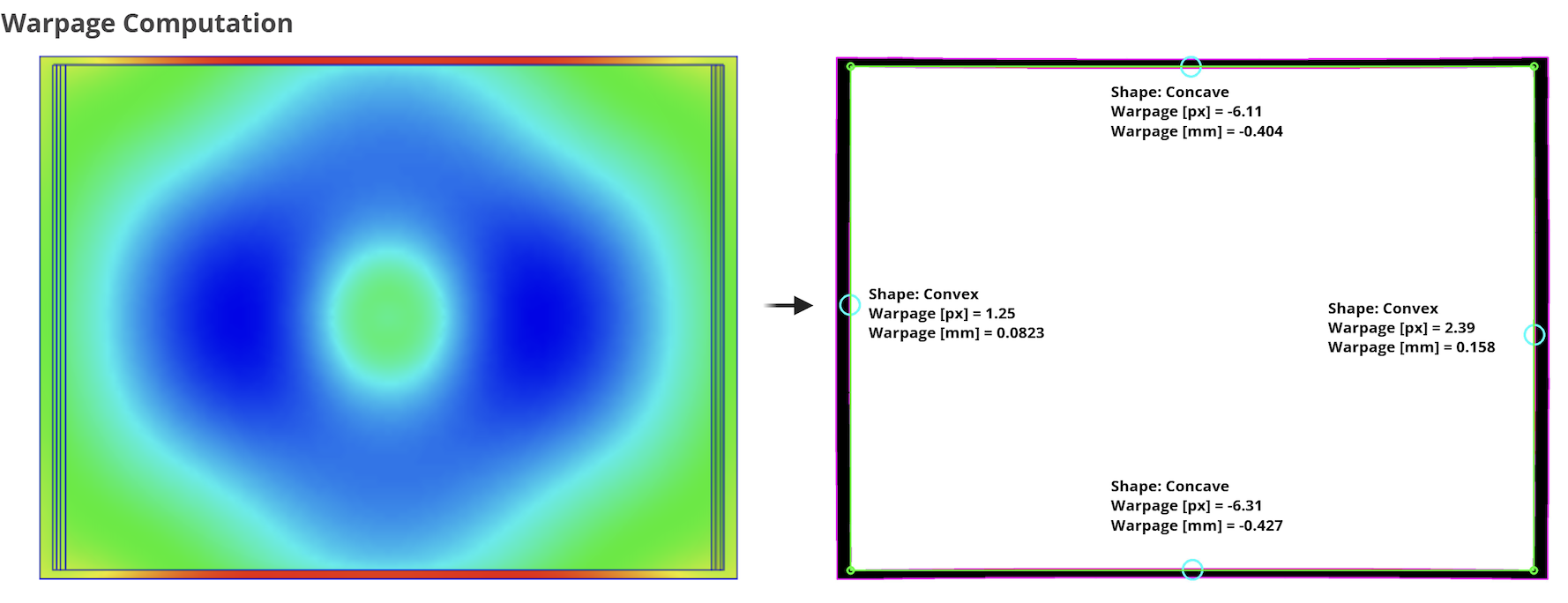}
    \caption{Post-processed 2D warpage contour images from Moldex3D are analyzed to convert pixels into deformation units i.e. mm.}
    \label{fig:Partdesign}
\end{figure}

We apply the proposed uncertainty analysis approach to the injection molding simulation illustrated in Figure~\ref{fig:Partdesignwarpage}. A total of 57 simulation runs were performed in the commercial injection molding software Moldex3D using 57 distinct combinations of the four input parameters. The transient cooling analysis with warpage module in Moldex3D was employed to capture the resulting deformation behavior. The warpage of the four designated wall sections (horizontal left, horizontal right, vertical up, and vertical down) was quantitatively evaluated using computer vision techniques, as illustrated in Figure~\ref{fig:Partdesign}. The ranges of all input and output variables are summarized in Table~\ref{tb:summary}.


\begin{table}[!h]
\centering
\caption{Summary of Input and Output Parameters of the Injection Molding Simulation}\label{tb:summary}
\begin{tabular}{lcc}
\toprule
\textbf{Parameter} & \textbf{Range} & \textbf{Unit} \\
\midrule
\multicolumn{3}{l}{\textit{Input Parameters}} \\
Mold Temperature  & 30--50  & $^\circ$C \\
Injection Speed   & 22.5--67.5 & mm/s \\
Packing Pressure  & 400--600 & MPa \\
Packing Time      & 1.0--4.5 & s \\
\midrule
\multicolumn{3}{l}{\textit{Output Responses}} \\
Horizontal Left   & $[-1.134,\ 1.632]$ & mm \\
Horizontal Right  & $[-1.087,\ 1.661]$ & mm \\
Vertical Up       & $[-0.971,\ 0.426]$ & mm \\
Vertical Down     & $[-1.046,\ 0.415]$ & mm \\
\bottomrule
\end{tabular}
\end{table}

Following the notation introduced in Section~\ref{sec2}, the four-dimensional input vector is denoted by 
\[
\bm{x} = (x_1, x_2, x_3, x_4)^\top,
\]
and the corresponding four-dimensional output vector is denoted by 
\begin{equation}\label{eq:ss}
\bm{y}(\bm{x}) = \big(y_1(\bm{x}),\, y_2(\bm{x}),\, y_3(\bm{x}),\, y_4(\bm{x})\big)^\top.
\end{equation}
The objectives of this study are twofold. 
First, we aim to identify the optimal parameter setting that minimizes the overall deformation across the four wall directions:
\[
\bm{x}^* \in \arg\min_{\bm{x} \in \mathcal{X}} \sum_{l=1}^4 y_l^2(\bm{x}),
\]
and to quantify the uncertainty associated with this optimal solution. 
Second, for each direction \(l = 1, 2, 3, 4\), we seek to locate the boundary in the input space separating positive and negative displacements, defined as
\[
\partial \mathcal{X}_l = \{\bm{x} \in \mathcal{X} \mid y_l(\bm{x}) = 0\},
\]
and to construct a confidence band around this boundary to characterize the associated uncertainty.

Following Section~\ref{sec2}, 
we fitted second-order polynomial regression models for each of the four response variables. 
Table~\ref{tab:R^2} presents the coefficient of determination ($R^2$) for each model. 
The results indicate that the fitted models provide accurate and reliable surrogates for the simulation outputs. Based on these models, the optimal decision as in \eqref{eq:surrogate} is given by 
$\hat{\bm{x}}^* = (43.256, 49.297, 437.282, 4.500)^\top$, corresponding to mold temperature, injection speed, packing pressure, and packing time, respectively. The resulting minimum sum of squared displacements is 0.00034 $\text{mm}^2$.
As illustrated in Sections \ref{sec2} and \ref{sec3}, the followup decision uncertainty estimation and boundary analysis are also based on these models.

\begin{table}[ht]
    \centering
    \caption{${R}^2$ for the Second-order Polynomial Regression Model from Each Output }
    \label{tab:R^2}
    \begin{tabular}{c|c|c|c}
        \toprule
       Horizontal Left $(y_1)$ & Horizontal Right $(y_2)$ & Vertical Up $(y_3)$ & Vertical Down $(y_4)$ \\
        \midrule
        $0.93$ & $0.94$ & $0.90$ & $0.93$ \\
        \bottomrule
    \end{tabular}
\end{table}

\begin{figure}[h]
  \centering
    \includegraphics[width=1\linewidth]{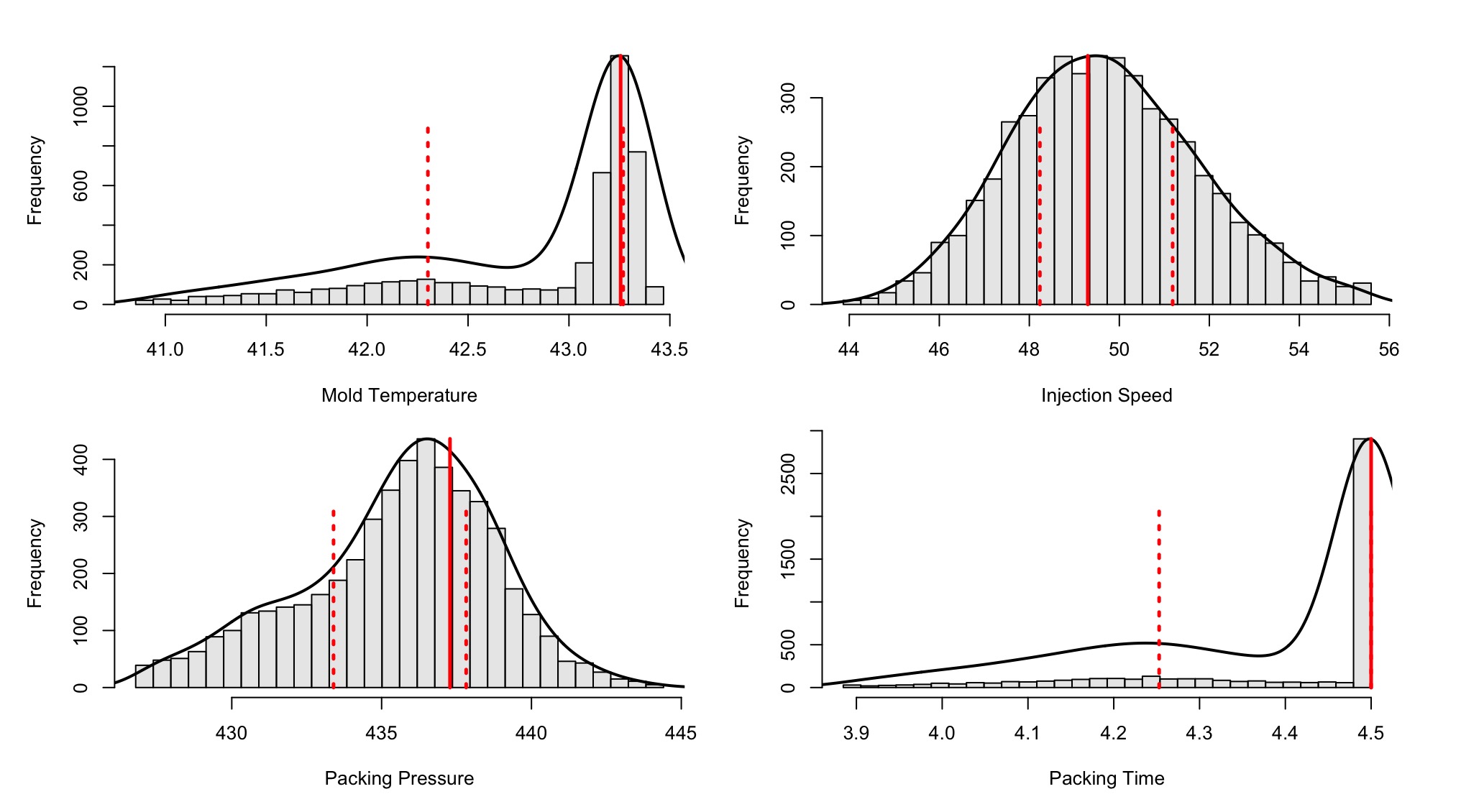}
  \caption{Marginal Distribution of the four inputs based on the decision uncertainty estimation procedure. The solid red line denotes the median of the optimal solutions, while the dashed lines indicate the 25th and 75th percentiles.}
  \label{fig:uq}
\end{figure}

\paragraph{Decision Uncertainty Estimation}
Following the decision uncertainty estimation procedure described in Section~\ref{sec2}, 
we generated 1,000 realizations of the fitted model and obtained the corresponding distributions of optimal solutions using the L-BFGS-B optimization algorithm \citep{byrd1995limited} to form a Monte Carlo sample of the distribution of the optimal decision. 
Figure~\ref{fig:uq} illustrates the marginal distributions of the optimal solutions in each input dimension based on Monte Carlo sampling. 
The solid red line denotes the median of the optimal solutions, while the dashed lines indicate the 25th and 75th percentiles, reflecting the variability around the median. 
By analyzing these distributions, we can quantify the uncertainty associated with each input variable and assess the robustness of the recommended optimal parameter settings under different realizations.

\begin{figure}[H]
    \centering
    \includegraphics[scale=0.6]{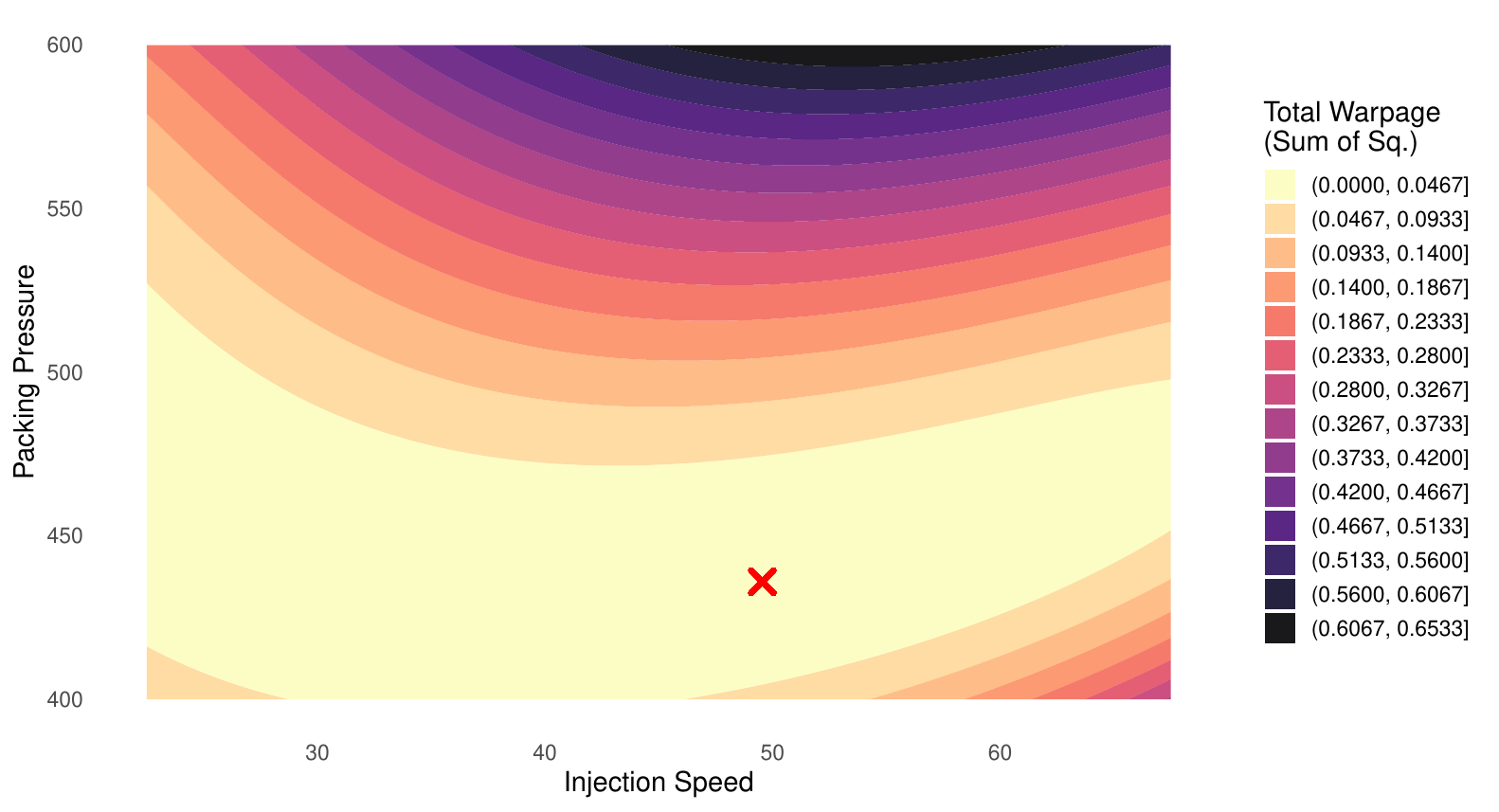}
    \caption{Contour plot of the predicted objective function in \eqref{eq:ss} as a function of injection speed and packing pressure. Mold temperature and packing time are fixed at their median optimal values. The red cross mark indicates the location of the median optimal solution, which aligns with the region of minimum warpage (the dark valley).}
    \label{fig:obj_contour}
\end{figure}

To visualize the behavior of the objective function, we examine the response surface of the total warpage. Figure \ref{fig:obj_contour} shows the contour plot of the sum of squared displacements in \eqref{eq:ss} with respect to injection speed and packing pressure. In this plot, mold temperature and packing time are fixed at their median optimal values from Figure \ref{fig:uq}. The color gradient represents the warpage level, where darker regions indicate lower values. The plot reveals a clear basin around the optimal region. This confirms that the objective function is stable and the identified solution lies in a region of minimal warpage. 

\paragraph{Boundary Analysis with Uncertainty}
To further quantify the uncertainty in our surrogate-based optimization for injection molding, we apply the boundary analysis method described in Section~\ref{sec3} to the simulation data. Specifically, we construct a Monte Carlo-based confidence band for the surrogate models of warpage outcomes, visualizing the regions where the predicted response is most uncertain.

The visualization of the boundary can only be performed in a two-dimensional input space. 
Therefore, in this analysis, we illustrate the results by fixing two of the four process variables (i.e., mold temperature and packing time) at their respective optimal settings, as shown in Figure~\ref{fig:uq}. 
The remaining two variables, injection speed and packing pressure, are varied to visualize the boundary behavior.
 Hereafter, we denote injection speed by $x_2$ and packing pressure by $x_3$. For each of the four responses, we use the fitted quadratic regression model to generate a dense grid over injection speed and packing pressure. At each grid point, we sample from the posterior distribution of the regression coefficients (as described in Section~\ref{sec2}), compute the predicted response, and standardize each prediction by its posterior standard deviation. By repeating this process for a large number of samples, we estimate the probability that the standardized response at each point satisfies $\lvert y_k(x_2,x_3)/\sigma_{y_k}(x_2,x_3)\rvert < \varepsilon$.

\begin{figure}
    \centering
    \includegraphics[scale=0.63]{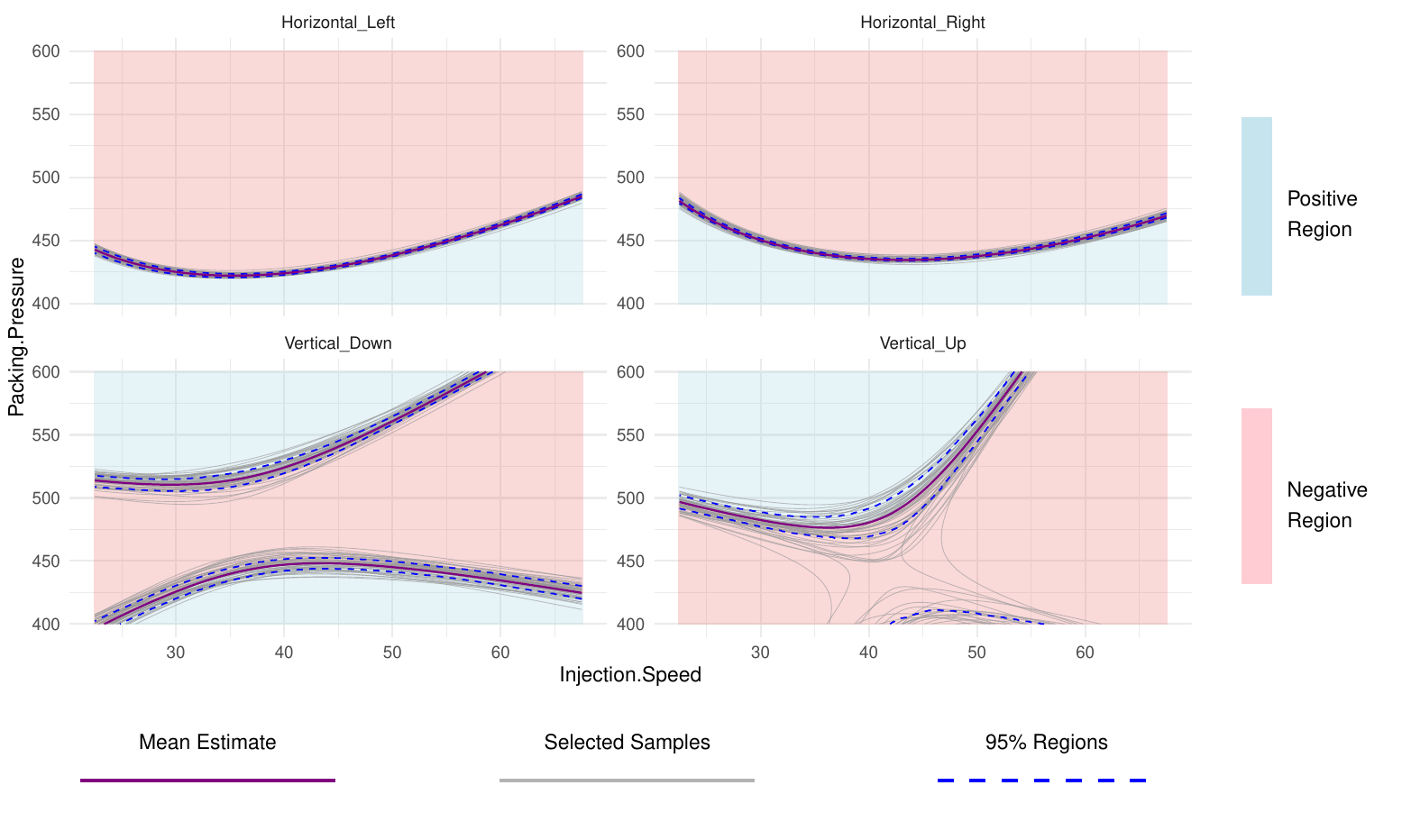}
    \caption{Monte Carlo–based 95\% confidence bands for Injection Molding Data using standardized tolerance. Each blue dashed contour shows the region where \(P\bigl(\lvert y_k(x_2,x_3)/\sigma_{y_k}(x_2,x_3)\rvert < \varepsilon\bigr)\ge0.95, \varepsilon=2.5\). The purple curve is the mean estimate zero-level set, and the gray curves are twenty sample zero-level sets from the posterior.}
    \label{fig:im-band}
\end{figure}

The resulting boundary plot in Figure~\ref{fig:im-band} displays the region where the empirical coverage probability
\[
\mathbb P\bigl(\lvert y_k(x_2,x_3)/\sigma_{y_k}(x_2,x_3)\rvert < \varepsilon\bigr) \ge 0.95.
\]
The blue dashed contour marks this 95\% confidence region. The purple solid curve represents the posterior mean zero-level set, and the gray curves denote twenty randomly selected posterior sample zero-level sets for illustration of boundary variability.  

\section{Conclusion}
\label{sec5}
In this study, we present a surrogate-based optimization framework to address the warpage problem in injection molding. Our primary goal is to minimize the dimensional deformation of a box-shaped part while explicitly accounting for parameter uncertainty. We use quadratic polynomial regression to approximate the complex relationship between process parameters and the resulting warpage.
The key contribution of this work lies in the integration of a Bayesian approach with polynomial surrogate models to characterize decision and boundary uncertainty. Our framework provides the uncertainty analysis of the optimal decision. This allows engineers to assess the variability and robustness of the recommended settings. Additionally, we introduce a boundary analysis technique using Monte Carlo simulation. This method successfully visualizes the confidence bands for the zero-warpage boundary, clearly distinguishing between convex and concave displacement regions. By addressing the mathematical challenges of inverting confidence bands analytically, the proposed simulation-based approach offers a practical tool to understand process limits.

\section*{Acknowledgment}
This work was supported as part of the AIM for Composites, an Energy Frontier Research Center funded by the U.S. Department of Energy, Office of Science, Basic Energy Sciences at the Clemson University under award \#DE-SC0023389.

\bibliographystyle{chicago}
\bibliography{ref}

\end{document}